\documentclass[conference]{IEEEtran}

\ifCLASSINFOpdf
   \usepackage[pdftex]{graphicx}
\else
   \usepackage[dvips]{graphicx}
\fi

\begin{document}

\title{Wet Paper Coding for Watermarking of Binary Images}

\author{\IEEEauthorblockN{Michail Zubarev and Valery Korzhik}
\IEEEauthorblockA{State University of Telecommunications\\ St. Petersburg\\ Russia\\ Email: korzhik@spb.lanck.net}
\and
\IEEEauthorblockN{Guillermo Morales-Luna}
\IEEEauthorblockA{Computer Science\\ CINVESTAV-IPN\\ Mexico City\\ Mexico\\ Email: gmorales@cs.cinvestav.mx}}

\maketitle

\begin{abstract}
We propose a new method to embed data in binary images, including scanned text, figures, and signatures. Our method relies on the concept of wet paper codes. The shuffling before embedding is used in order to equalize irregular embedding capacity from diverse areas in the image. The hidden data can be extracted without the original binary image. We illustrate some examples of watermarked binary images after wet paper coding.
\end{abstract}

\IEEEpeerreviewmaketitle

\section{Introduction}

A great majority of works on image watermarking (WM) has been devoted to color or gray scale images where pixel luminances may take values on a wide range. Then slight changings of even most of pixel luminance values may not be noticeable under normal viewing conditions.

Data hiding on digital binary images (BI) has been investigated in~\cite{one}. The main characteristic of this approach is the notion of {\em flippable pixels} which are precisely those that can be inverted (flipped) without noticeable corruption of BI. In our scheme, a bit string is embedded in each block based on the flippability score of the pixel within the block. This score is measured by the changes in smoothness and connectivity. A shuffling table or a key is required for correct extraction of the bit string watermark. In~\cite{two}, Wu {\em et al.} have presented a flippability score computation method by considering the smoothness and connectivity of the cover image. In their work, the score is determined by considering the changes in smoothness and connectivity and it is reached by the proposed rules. However, the involved analysis is quite extensive for neighborhoods larger than $3\times 3$ pixels. Another flippability score computation method was proposed in~\cite{three}. That method is based on the reciprocal distance calculation, which matches well within subjective evaluation, and it is easy to extend to block neighborhoods larger than $3\times 3$ pixels. In~\cite{four}, the authors use a weight matrix to calculate the new transitions created by the flipping. That approach calculates the distortion by combining the changes in smoothness and in connectivity. Maxemchuk {\em et al.}~\cite{five} changed line spacing and character spacing in order to embed information in textual images for bulk electronic publications. This method cannot be easily extended to other binary images and the size of hidden data is very small.

In all those approaches the embedding capacity is not large. In order to deal with this limitation, an approach has been proposed, called {\em writing on wet paper}~\cite{six}.

In our proposed procedure, we exploit the notion of flippable pixels, as presented in~\cite{one}, and we combine it with the {\em wet paper codes} (WPC) given in~\cite{six}. In section 2 we show a method for finding flippable pixels. Section 3 contains the description of the WPC technique. Section 4 contains some experimental results. Section 5 contains some conclusions and poses resulting open problems.

\section{WM Embedding Method for BI}

It was mentioned already that it is rather difficult to embed a WM into a BI without noticeable distortion. In~\cite{one} an approach has been developed to look through each of the adjacent \mbox{$N\times N$} areas of pixels and to embed a ``0'' in an area by either changing or not one of its pixels, so that the total number of black pixels in the area is an even number; similarly, to embed a ``1'', by either changing or not one pixel, so that the number of black pixels is an odd number. If it is necessary to change one pixel in an area it has to be chosen as a {\em flippable pixel}. Obviously the area should be non-empty. We will specify flippable pixels in greater detail.

\begin{figure}
\centering
\includegraphics[width=3.5in]{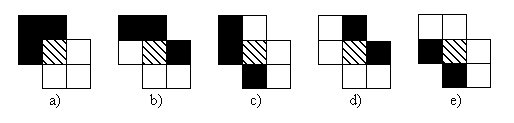} 
\caption{$3\times 3$ patterns with flippable center pixels.}
\label{fg.01}
\end{figure}

Let us consider initially $3\times 3$ areas with the intention to change only the central pixel (whenever it was necessary). In Figure~\ref{fg.01} we present the patterns in which we will consider the central pixel as flippable, of course the central pixel of any pattern obtained by a rotation around the central axis or an inversion of those patterns is considered also as flippable.
We remark that our choice of {\em central-flippable patterns} (CFP) is of a heuristic kind because we are mainly concerned in maintaining the connectivity features after a flipping. We may try to consider a larger area size, $N>3$, but this will involve the same $3\times 3$ patterns displayed at Figure~\ref{fg.01}. Namely, we will shift a $3\times 3$ area, by one pixel in turn, in both horizontal and vertical directions, along the $N\times N$ area until all flippable pixels are found.

\section{Writing of Wet Paper for the Case of Binary Images}

This name derives from a real situation, when the paper has been exposed to rain. The sender is able to write the message only over dry places. During transmission the paper dries out and the recipient does not know, which places were dry when encoding. However he is able to read the message easily. In information theory this technique is known as {\em ``writing in memory with defective cells''}.

\subsection{Encoder}

Let us assume that the sender has a binary column vector ${\bf b}=\left(b_i\right)_{i=1}^n$ received from a binary image of $n$ pixels (a black pixel gives value 1, and a white pixel gives 0), and an index set $C\subset\{1,\ldots,n\}$, $|C|=k$, corresponding to those bits that are flippable and subject to be modified in order to embed a message. The sender wants to communicate a sequence of $q$ bits, ${\bf m}=(m_1,\ldots,m_q)$. First, let us assume that the recipient knows $q$ and let us postpone later the case when the recipient does not know $q$. The sender and recipient use a shared stegokey to generate a pseudo-random binary $(q\times n)$-matrix $D$. The sender shall modify some flippable bits $b_i$, $i\in C$, in such a way that the modified binary column vector ${\bf b}'=\left(b_i'\right)_{i=1}^n$ satisfies $D{\bf b}' = {\bf m}$. Let ${\bf v} = {\bf b}'-{\bf b}$, hence the support of vector ${\bf v}$ is a subset of $C$, then 
\begin{equation}
D{\bf v} = {\bf m} - D{\bf b} \label{eq.02}
\end{equation}
Given ${\bf m}$ and ${\bf b}$, the system of linear equations~(\ref{eq.02}) should be solved for $k$ unknowns $v_i$, $i\in C$ (clearly, $v_i=0$ whenever $i\not\in C$). By deleting the rows and columns with indexes in the complement of $C$, the system~(\ref{eq.02}) can be rewritten as
\begin{equation}
H{\bf v}' = {\bf m} - D{\bf b} \label{eq.03}
\end{equation}
with $H$ of order $(q\times k)$ and ${\bf v}'=\left(v_i\right)_{i\in C}$ a $k$-dimensional vector. The equations system~(\ref{eq.03}) is posed in the Galois field $\mbox{GF}(2)$ and it has an unique solution whenever $\mbox{rank}(H)=q$. Thus, in order to decide which flippable pixels should actually be flipped, the encoder should solve the equations system~(\ref{eq.03}).

\subsection{Decoder}

The modified binary image ${\bf b}'$ is sent to the recipient. The decoding is very simple because the recipient may recover the message ${\bf m}=D{\bf b}'$ using the shared matrix $D$. Note that the recipient is not required to know which pixels are flippable.

We now explain how to relax the assumption that the recipient knows $q$. The sender and the recipient can agree matrix $D$ in a row-by-row manner rather than as a whole planar $(q\times n)$-array of bits. The sender may reserve the first $\lceil\log_2 n\rceil$ bits of the message ${\bf m}$ as a header to inform the recipient of the number of rows in $D$ (the symbol $\lceil x\rceil$ denotes the smallest integer greater or equal than $x$). The recipient takes the first $\lceil\log_2 n\rceil$ rows of $D$, and applies the resulting $(\lceil\log_2 n\rceil\times n)$-submatrix to the received vector ${\bf b}'$, in order to recover the header, i.e. the message length $q$. Then, he uses the rest of $D$ to recover the message ${\bf m}=D{\bf b}'$.

\subsection{Practical implementation}

Due to the great computational complexity of solving a system~(\ref{eq.03}) for a big value of $q$, we partition our image into areas of $N$ pixels, and for each we solve a corresponding system~(\ref{eq.03}) of linear equations. Besides, since there is no a regular distribution of flippable pixels from area to area, we use shuffling of the image pixels as it was suggested in~\cite{seven}.

If the length of the embedded message is greater than the number of flippable pixels in the area, we try to embed as many bits as possible. The remaining bits would be embedded into following areas. As remarked before, whenever condition $\mbox{rank}(H)=q$ is fulfilled for an area, we may embed \mbox{$k-\lceil\log_2 n\rceil$} bits in that area. Otherwise, we may decrease $q$ until condition $\mbox{rank}(H)=q$ is satisfied.

\section{Experimental results}

We took $n=4096$ pixels as the area size. In Table~\ref{tb.01} we present the numbers $N_A$, $N_{FP}$, $N_E$ corresponding to image areas, flippable pixels in an image and embedded bits for three typical images. 
\begin{table}
\begin{center}
 \begin{tabular}{||l|r|r|r||}      \hline \hline
         & Image 1 & Image 2 & Image 3 \\ \hline
$N_A$    &      16 &      16 &      16 \\ \hline
$N_{FP}$ &    1283 &    5455 &    2358 \\ \hline
$N_E$    &    1058 &    5222 &    2122 \\ \hline
\hline
\end{tabular}
\end{center}
 \caption{The numbers of image areas $N_A$, flippable pixels in the image $N_{FP}$, and embedded bits into an image $N_E$.\label{tb.01}}
\end{table}

\begin{figure*}
\centering
\begin{tabular}{cc}
 \includegraphics[width=2in]{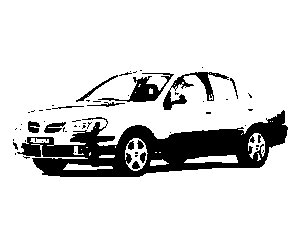} & \includegraphics[width=2in]{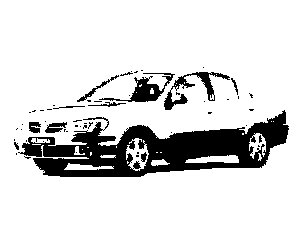} \\
(a) & (b) \\
\multicolumn{2}{c}{Image 1} \vspace{2ex} \\
 \includegraphics[width=2in]{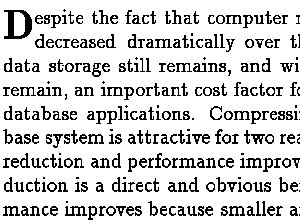} & \includegraphics[width=2in]{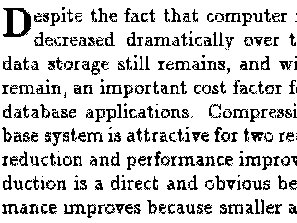} \\
(a) & (b) \\
\multicolumn{2}{c}{Image 2} \vspace{2ex} \\
 \includegraphics[width=2in]{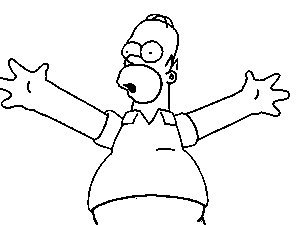} & \includegraphics[width=2in]{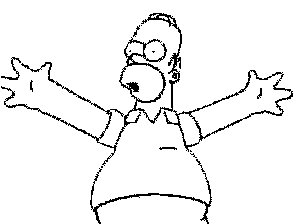} \\
(a) & (b) \\
\multicolumn{2}{c}{Image 3} 
\end{tabular}
\caption{Images 1, 2, 3 in (a) original and (b) after watermarking with WPC.}
\label{fg.02}
\end{figure*}

The three processed images, with actual size of about $300\times 225$ pixels, are presented in their original forms (before watermarking) and after watermarking in Figure~\ref{fg.02}.

\section{Conclusions}

It seems to be difficult the embedding into a BI of an additional information without remarkable distortions of the original image. But as it can be appreciated in Table~\ref{tb.01} and Figure~\ref{fg.02}, it is indeed possible to embed a significant number of information bits without any visual difference among the original images and their watermarked copies. 

It is an interesting open problem to find an efficient statistical method in order to distinguish the original BI from the watermarked BI. Solutions to this steganography problem are well known~\cite{xii,xiii} for gray-scale (or color) images but they have a poor adequacy for BI. Another interesting open problem is to develop a technique for {\em selective authentication} of BI when there is a group of legitimate and illegitimate distortions of the originals. If it occurs that distortions belong to the first group, then authentication is confirmed, while if it occurs at the second group, then authentication fails. The selective authentication is especially important for BI because the legitimate distortion can be resulted from channel noises or corruption after scanning of paper documents or images received by fax. We are going to present our contribution in this direction in the near future.

\section*{Acknowledgment}

Dr. Morales-Luna acknowledges the support of Mexican Conacyt.

\bibliographystyle{IEEEtran}
\bibliography{08zui}

\end{document}